\documentclass[12pt,preprint]{aastex}

\shorttitle{Spectrum Analysis of SN 1996cb}

\shortauthors{Deng, Qiu \& Hu}

\begin{document}

\title{Spectrum Analysis of Type IIb Supernova 1996cb}

\author{Jinsong~Deng \altaffilmark{1,2}}

\affil{Research Center for the Early Universe, School of Science,
University of Tokyo, Bunkyo-ku, Tokyo 113-0033, Japan}

\email{deng@astron.s.u-tokyo.ac.jp}

\author{Yulei~Qiu and Jinyao~Hu}

\affil{Beijing Astronomical Observatory, Chaoyang District,
Beijing 100012, P.R.China}

\email{qiuyl@nova.bao.ac.cn; hjy@class1.bao.ac.cn}

\altaffiltext{1}{Department of Astronomy, School of Science,
University of Tokyo, Bunkyo-ku, Tokyo 113-0033, Japan}

\altaffiltext{2}{Beijing Astronomical Observatory, Chaoyang
District, Beijing 100012, P.R.China}

\begin{abstract}

We analyze a time series of optical spectra of SN 1993J-like
supernova 1996cb, from 14 days before maximum to 86 days after
that, with a parameterized supernova synthetic-spectrum code
SYNOW. Detailed line identification are made through fitting the
synthetic spectra to observed ones. The derived photospheric
velocity, decreasing from $11,000{\rm~km~s^{-1}}$ to $3,000{\rm~
km~s^{-1}}$, gives a rough estimate of the ratio of explosion
kinetic energy to ejecta mass, i.e. $E/M_{\rm ej}\sim
0.2-0.5\times 10^{51}{\rm~ergs}/M_{\rm ej}(M_{\sun})$. We find
that the minimum velocity of hydrogen is $\sim
10,000{\rm~km~s^{-1}}$, which suggests a small hydrogen envelope
mass of $\sim 0.02-0.1~M_{\rm ej}$, or $0.1-0.2~M_\sun$ if $E$ is
assumed $1\times10^{51}{\rm~ergs}$. A possible \ion{Ni}{2}
absorption feature near $4000{\rm~\AA}$ is identified throughout
the epochs studied here and is most likely produced by primordial
nickel. Unambiguous \ion{Co}{2} features emerge from 16 days after
maximum onward, which suggests that freshly synthesized
radioactive material has been mixed outward to a velocity of at
least $7,000{\rm~km~s^{-1}}$ as a result of hydrodynamical
instabilities. Although our synthetic spectra show that the bulk
of the blueshift of [\ion{O}{1}] $\lambda 5577$ net emission, as
large as $\sim70{\rm~\AA}$ at 9 days after maximum, is attributed
to line blending, a still considerable residual $\sim20{\rm~\AA}$
remains till the late phase. It may be evidence of clumpy or
large-scale asymmetric nature of oxygen emission region.

\end{abstract}


\section{Introduction}

SN 1996cb in NGC 3510 was first discovered by Aoki, Cho, and
Toyama on 1996 December 15 \citep{nak96}, and independently by
BAOSS on 1996 December 18 \citep{qia96}. It was classified as a
type IIb event due to spectral similarity to SN 1993J
\citep{gar97}. As reviewed by \citet{fil97}, this subclass of SNe
II appear to be links between normal SNe II and SNe Ib. They show
strong hydrogen Balmer lines near maximum brightness, then evolve
with the disappearance of hydrogen features and the emergence of
helium ones. Many days after maximum, their spectra are dominated
by oxygen and calcium emission lines and hence resemble those of
SNe Ib at the nebular phase. Their progenitors are usually thought
to be massive stars, having lost most of their hydrogen envelopes
either through stellar winds \citep[e.g.][]{hof93}, or more likely
as the results of mass exchange with their companions
\citep[e.g.][]{nom93,pod93}.

Very few type IIb events have been recognized to date. The first
one is SN 1987K \citep{fil88}, unfortunately whose proximity to
the Sun made observations unavailable for a long period after
maximum. Unlike it, the famous bright SN 1993J has been exposed to
intensive observations and theoretical analysis \citep[for a brief
review, see][]{mat00a}. For SNe 1997dd and 1998fa, it is only very
recently that one and four spectra respectively were published
\citep{mat01}. No data of SNe 1996B \citep{wan96}, 2000H
\citep{ben00}, 2001ad \citep{cho01}, and 2001cf \citep{fil01} have
been released yet. Other proposed candidates, like SNe 1989O
\citep{fil89} and 1999bv \citep{hil99}, were either distant or
discovered at the late phase, with their classifications some
uncertain.

\citet{qiu99} have published $BVR_{c}$ light curves of SN 1996cb
and its low-resolution optical spectra, obtained at Beijing
Astronomical Observatory, from 14 days before maximum to 160 days
after that. The light curves are very similar to those of SN
1993J, except that in SN 1996cb the initial peak due to shock
breakout and the subsequent rapid decline have not been observed.
The $B$ light curve reached a maximum of 14.22 mag on 1997 January
2, which makes it the third brightest supernova discovered in
1996. By comparing its \bv color curve with that of SN 1993J, they
estimated the explosion date at 1996 December 12. They also
described the observed spectral evolution in a brief way, while
emphasizing the blueshift of \ion{He}{1} $\lambda 5876$ and
[\ion{O}{1}] $\lambda 5577$ emission peaks. In SN 1993J, such a
phenomenon for [\ion{O}{1}] $\lambda 5577$ has been suggested
being evidence of Rayleigh-Taylor instabilities at the He/C+O
interface \citep{wan94}.

It is important to analyze and model spectroscopic and photometric
data of SN 1996cb, by far the only other well-observed SN
1993J-like supernova available for comprehensive investigations.
The results can be compared with those of SN 1993J. Numerous
studies on the latter arrived at the conclusion that it is the
explosion of a massive star with a $\sim4~M_\sun$ helium core and
a small He-rich hydrogen envelope of $\sim0.1-0.9~M_\sun$
\citep[etc]{nom93,pod93,shi94,bar94b,utr94,woo94,you95}. These
studies also have great influences on many aspects of our
understanding of core-collapse supernovae, like the pre-supernova
evolution in a binary system \citep[e.g.][]{nom95}, asymmetric
nature of explosion \citep[e.g.][]{wan01}, large-scale instability
and mixing \citep[e.g.][]{iwa97}, NLTE and non-thermal effects on
spectrum formation \citep[e.g.][]{hou96}, interaction of ejecta
with circumstellar matter \citep[e.g][]{fra96}, etc.

In this paper, we report our work on the synthesis of a spectra
series of SN 1996cb from 1996 December 19 to 1997 March 28,
spanning almost its whole photospheric stage. We use a
parameterized supernova synthetic-spectrum code SYNOW. The method
and parameters are described in \S2. Our main goal is to establish
line identifications, which are presented in \S3 with our best fit
synthetic spectra. Discussions on photospheric velocity, velocity
and mass of hydrogen envelope, identifications of \ion{Ni}{2} and
\ion{Co}{2}, blueshift of [\ion{O}{1}] $\lambda 5577$, etc. are
given in \S4. The comparison with SN 1993J is mentioned wherever
within reach. Our main conclusions are summarized in \S5.

\section{Spectrum Synthesis Procedure}

To establish reliable line identifications, theoretical synthesis
of supernova spectra is necessary. SYNOW is a parameterized
supernova synthetic-spectrum code developed by Branch and
collaborators and especially suitable for such a purpose
\citep{bra80, jef90, fis00}. The current version of SYNOW uses 42
million line list of \citet{kur93}. It can treat the complicated
line blending properly, but still retain a low cost of run-time.
We know line blending is a great nuisance in the analysis of
supernova spectra, which corresponds to nonlocal radiative
coupling between different lines due to a large spatial gradient
of velocity \citep{ryb78,ols82}. Numerical tests show that in the
blending of a group of lines even features with large optical
depth can be greatly suppressed and sometimes a single
quasi-P~Cygni profile, but unusually shaped, will be yielded
\citep{jef90}.

The basic physical assumptions and parameters adopted in SYNOW can
be described as follows: The supernova ejecta is assumed
spherically symmetric and with radial velocity $v\propto r$ at any
radius $r$, namely in homologous expansion. The observed continuum
is assumed to arise out of a sharp photosphere and fitted by a
blackbody one with temperature $T_{\rm bb}$, while formation of
spectral lines are treated as resonant scatterings of continuum
photons above the photosphere in Sobolev approximation. No
detailed ionization and excitation balance is considered; hence
for each ion we choose the Sobolev optical depth, at the
photosphere, of one reference line as a free parameter, with those
of the others fixed by LTE assumption. The radial dependence of
line optical depths is assumed to follow the density profile of
ejecta, which is simplified to be either a power-law one,
$\rho\propto\exp(-v/v_e)$, or an exponential one,
$\rho\propto\exp(-v/v_e)$. The velocity of photosphere $v_{\rm
ph}$ is an important fitting parameter.

In this work, we adopt an exponential density profile,
$\rho\propto\exp(-v/v_e)$, with e-folding velocity $v_e=2,000
{\rm~km~s^{-1}}$. To determine which ions should be invoked when
computing a synthetic spectrum, we refer to line identifications
for other Type II and Ib/c supernova, especially SN 1987A
\citep[e.g][]{jef90} and SN 1993J \citep[e.g.][]{mat00a}. The
systematic plots of LTE Sobolev line depths versus temperature,
made by \citet{hat99a}, for some common supernova compositions are
also useful. For each ion, the LTE excitation temperature $T_{\rm
exc}$ is also a fitting parameter but bears little physical
significance. We adapt $T_{\rm exc}$ in the range from
$4,000{\rm~K}$ to $12,000{\rm~K}$. We fix the outer edge of the
line-forming region at $50,000{\rm~km~s^{-1}}$. Sometimes we have
to set for an ion a minimum velocity $v_{\rm min}$, greater than
$v_{\rm ph}$, which shows that it is detached from the
photosphere.

To make exact comparisons with synthetic spectra, the observed
spectra should be transformed to the rest frame of the host galaxy
and corrected for possible interstellar reddening. The recession
velocity of NGC 3510 adopted here is $\sim770{\rm~km~s^{-1}}$
\citep{gar97}. By comparing \bv color curves, \citet{qiu99} argued
that the color excess of SN 1996cb is 0.2 lower than that of SN
1993J. Although various values were found for the latter, the
average seems $\sim0.2$ \citep[see][and references
therein]{mat00b}. So we assume $E(\bv)\approx0$ for SN 1996cb,
which is consistent with the negligible Galactic component
$\sim0.03$ \citep{sch98} and the absence of narrow \ion{Na}{1} D
absorption in a $50{\rm~\AA~mm^{-1}}$ dispersion spectrum taken on
1996 December 23 \citep{qiu99}.

\section{Spectrum Evolution and Line Identification}

\begin{figure}
\plotone{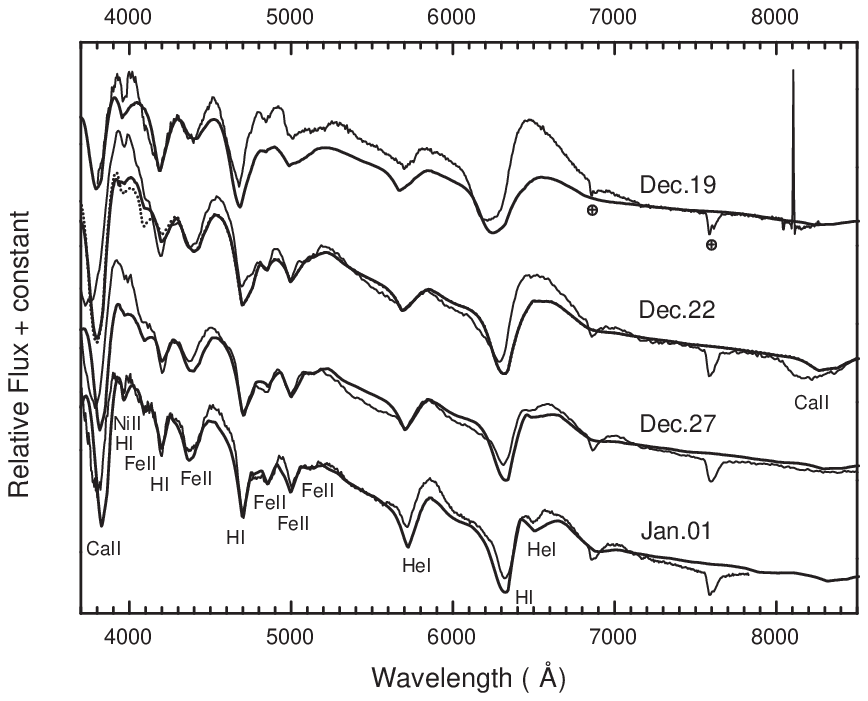} \caption{Observed spectra (thin solid lines) from
day 7 to day 20 compared with best fit synthetic spectra (thick
solid lines). \label{fig1}}
\end{figure}

\begin{figure}
\plotone{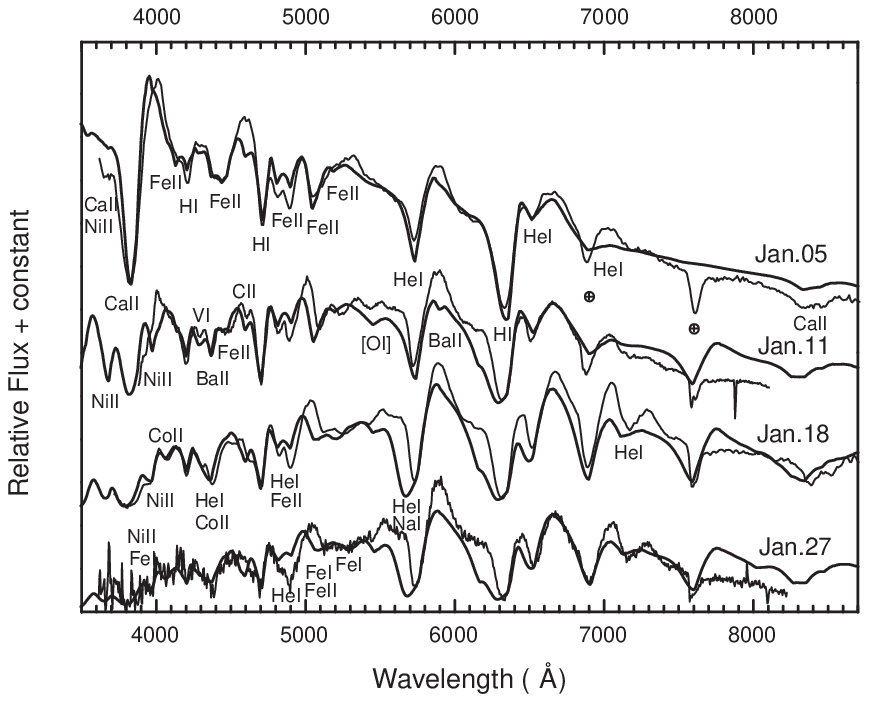} \caption{Observed spectra (thin solid lines) from
day 21 to day 46 compared with best fit synthetic spectra (thick
solid lines). \label{fig2}}
\end{figure}

For comparison with each observed spectrum, we have calculated
many synthetic spectra with various values of the fitting
parameters mentioned above. Our best fit synthetic ones are
plotted as thick solid lines in Figures \ref{fig1}, \ref{fig2},
\ref{fig3}, and \ref{fig4} together with observed ones (thin solid
lines) in time sequence. The telluric O$_2$ absorptions of A-band
($\sim7600{\rm~\AA}$) and B-band ($\sim6850{\rm~\AA}$) are marked.
We list in Table \ref{tbl1} the date of explosion, epoch, and best
fit value of $v_{\rm ph}$ and $T_{\rm bb}$. For each spectrum, two
different epoch denotations are employed in this paper. One is in
days after the estimated explosion date, 1996 December 12; and the
other is in days with respect to the date of B maximum, 1997
January 2, and always prefixed either a minus or a plus. For
example, our first spectrum, taken on 1996 December 19, can be
denoted as day 7 and -14 days respectively.

\subsection{Day 7 to 20}

The day 7 (-14 days) spectrum, shown in Figure \ref{fig1}, is
similar to those of SN 1993J from 7 to 23 days after explosion,
except the puzzling flat-topped characteristic of H$_\alpha$ in
the latter \citep[see][etc]{lew94,pra95,bar95b,fin95}. P~Cygni
profiles of H$_\alpha$, H$_\beta$, H$_\gamma$ are distinct.
\ion{Ca}{2} H\&K is strong, while the trough attributable to
\ion{Ca}{2} near-infrared triplet absorption is shallow. According
to our synthetic spectrum, P~Cygni features around $4400{\rm~\AA}$
and $4900{\rm~\AA}$, and the very broad triangular one from
$5000{\rm~\AA}$ and $5600{\rm~\AA}$ are produced by \ion{Fe}{2}
line blends. We can not reproduce the observed profile of
H$_\alpha$, which shows a net emission component with a
blueshifted peak and a rather broad absorption base. To fit a
prominent notch superimposed on the \ion{Ca}{2} H\&K emission, we
introduce \ion{Ni}{2} lines and blend them with H$_\delta$. The
strong P~Cygni feature around $5800{\rm~\AA}$ is identified as
\ion{He}{1} $\lambda5876$. The contribution from \ion{Na}{1} D is
supposed to be negligible because the temperature at such an early
phase, $\sim10,000{\rm~K}$, is high enough to ionize all neutral
sodium.

From day 10 (-11 days) to day 20 (-1 day), the top of H$_\alpha$
emission becomes flat and a notch gradually develops on it. To
simulate this behavior, we must fix $v_{\rm min}^{\rm HI}$ at
$10,500{\rm~km~s^{-1}}$, while the photospheric velocity decreases
from $10,000{\rm~km~s^{-1}}$ to $8,000{\rm~km~s^{-1}}$. It also
requires a drop in the relative optical depth, $\tau_{\rm
HI}/\tau_{\rm HeI}$, by one order from day 7 to day 20; and
accordingly the notch is identified as \ion{He}{1} $\lambda6678$.
In contrast to the first spectrum, here H$_\alpha$ profiles can be
well fitted under our LTE purely resonant scattering assumption,
which suggests that the hydrogen region is nearly recombined. All
these show that the photosphere has receded through the H/He
interface. In SN 1993J, this does not happen till 26 days after
explosion \citep[and references therein]{mat00a}. A new feature at
$\sim4100{\rm~\AA}$, produced by \ion{Fe}{2}, emerges from behind
the weakened H$_\delta$ absorption due to de-blending. The
``$4000{\rm~\AA}$'' notch is also attributed much more to
\ion{Ni}{2} than to H$_\delta$. Other line identifications are the
same as those on day 7. However, to fit the observed spectra, here
we must set $v_{\rm min}=10,500{\rm~km~s^{-1}}$ for \ion{Fe}{2}.

\subsection{Day 24 to 46}

Figure \ref{fig2} shows the spectra from day 24 (+3 days) to day
46 (+25 days). The day 24 spectrum resemble those previous, but
the constraint on $v_{\rm min}^{\rm FeII}$ is relaxed here and
after. \ion{He}{1} $\lambda7065$ develops clear. \ion{Ni}{2} helps
somewhat to form a feature around $3700{\rm~\AA}$ with
\ion{Ca}{2}, although the ``$4000{\rm~\AA}$'' notch that we
attribute to \ion{Ni}{2} and H$_\delta$ before disappears.

There are some noticeable changes in the day 30 spectrum: (1) The
broad single ``$5000-5600{\rm~\AA}$'' feature splits into three
distinct bumps. The blueward two are most likely produced by
\ion{Fe}{2} lines in strongly blending. There are some controversy
about the nature of the $\sim5500{\rm~\AA}$ feature in spectra of
SN 1993J, which can be traced back to as early as day 25
\citep{pra95,wan94}. Following \citet{wan94} and other authors, we
suggest it as the blueshifted [\ion{O}{1}] $\lambda5577$ emission.
By employing a low enough temperature of $T_{\rm exc}^{\rm
OI}=4,500{\rm~K}$, we can include forbidden lines [\ion{O}{1}]
$\lambda5577$ and $\lambda\lambda6300,6364$ in our synthetic
spectra, without causing an unreasonably strong \ion{O}{1}
$\lambda7773$. (2) The deep notch at $4370{\rm~\AA}$ and an
inconspicuous dip on the top of \ion{He}{1} $\lambda5876$ emission
can be fitted by \ion{Ba}{2} $\lambda6142$ and $\lambda4554$
respectively, if $v_{\rm min}^{\rm BaII}=13,000{\rm~km~s^{-1}}$ is
set. The dip just blueward to H$_\beta$ is identified as
$\lambda\lambda4738,4744$ absorption of \ion{C}{2} detached at
$v_{\rm min}=9,800{\rm~km~s^{-1}}$. We attribute the feature
located at $4290{\rm~\AA}$ to \ion{V}{1} absorption. (3)
\ion{Ca}{2} H\&K emission becomes weak and shows an unusual
triangular profile centered at $4006{\rm~\AA}$. The redward
decline can be well reproduced. But the fit to the blueward
incline, which is remarkably redshifted, can only be partially
improved by including \ion{Ni}{2} absorption.

Day 37 and day 46 spectra are dominated by \ion{He}{1}
$\lambda5876$ (blended with \ion{Na}{1} D), $\lambda6678$,
$\lambda7065$, and $\lambda7281$ P~Cygni lines in the red.
\ion{He}{1} also contributes a lot to absorptions at
$4380{\rm~\AA}$, $4820{\rm~\AA}$, and $4900{\rm~\AA}$. A distinct
notch appears at $4080{\rm~\AA}$ is identified as \ion{Co}{2}
feature through spectral synthesis. If \ion{Co}{2} is removed,
there will be obvious discrepancies also in other regions between
model spectra and observations. We noticed there is an absorption
feature near $4100{\rm~\AA}$ in the spectra of SN 1993J from day
56 to 109 \citep{mat00a,lew94,swa93b} but unfortunately without
identification. \ion{Ni}{2} $\sim4000{\rm~\AA}$ feature, with a
minor contribution from \ion{He}{1}, lies just redward to
\ion{Ca}{2} H\&K absorption, which becomes rather weak now. We
must decrease the optical depth of \ion{Fe}{2} to explain the
nearly absence of \ion{Fe}{2} $\lambda5169$. From day 46 onward,
\ion{Fe}{1} is introduced to lower the flux bluer than
$4500{\rm~\AA}$ to the observed level, and also helps to form the
basin-like shape between $5000{\rm~\AA}$ and $5500{\rm~\AA}$.
\ion{Ba}{2} and \ion{V}{1} features seen in day 30 spectrum have
completely disappeared.

\subsection{Day 51 to 107}

\begin{figure}
\plotone{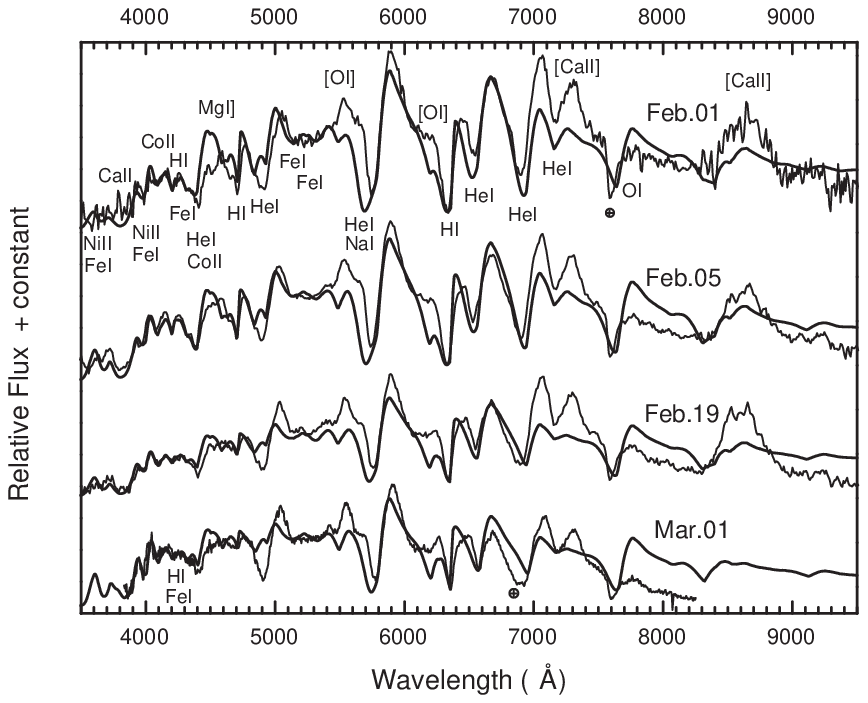} \caption{Observed spectra (thin solid lines) from
day 51 to day 80 compared with best fit synthetic spectra (thick
solid lines). \label{fig3}}
\end{figure}

\begin{figure}
\plotone{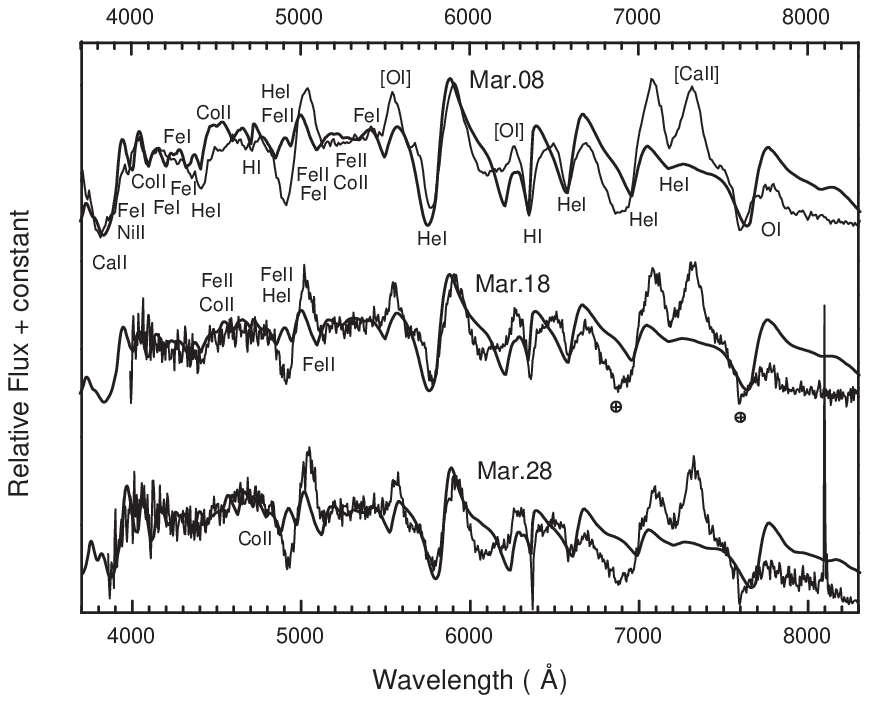} \caption{Observed spectra (thin solid lines) from
day 87 to day 107 compared with best fit synthetic spectra (thick
solid lines). \label{fig4}}
\end{figure}

Figures \ref{fig3} and \ref{fig4} show the spectra from day 51
(+30 days) to day 107 (+86 days). H Balmer lines continue to
weaken fast, consequently the spectra resemble those of Type Ib
more and more. Although a narrow H$_\alpha$ absorption and an
inconspicuous H$_\beta$ feature remain till day 107, they nearly
vanish in the April 4 spectrum (day 114) of \citet{qiu99} and
April 16 spectrum of \citet{mat01}. \ion{He}{1} line intensity
culminates around day 55, and then begins to decay, especially
that of $\lambda6678$. Because of this and also of the decrease of
photospheric velocity, telluric O$_2$ B-band ($\sim6850{\rm~\AA}$)
reappearances and contributes to a somewhat misleading
flat-bottomed profile.

Forbidden emissions like [\ion{Ca}{2}] $\lambda\lambda7291,7324$,
near-infrared triplet and narrow \ion{Mg}{1}] $\lambda4571$
develop in this period, which designates a transition to the
nebular phase. In accordance, the light curve of SN 1996cb enters
its tail around day 50 \citep{qiu99}. We can not produce these
nebular features with SYNOW due to their non-resonant-scattering
nature. [\ion{O}{1}] $\lambda5577$ does not strengthen much
throughout this epoch, while the gradually standing-out of
[\ion{O}{1}] $\lambda\lambda6300,6364$ emission is attributed more
to the weakening of H$_\alpha$ absorption than to the change of
oxygen abundance. There seems some \ion{O}{1} $\lambda7773$
emission, although the possible absorption is greatly contaminated
by telluric O$_2$ A-band.

\ion{Fe}{2} is removed from our synthetic spectra since day 51 and
is introduced again since day 87, in order to simulate the
evolution of \ion{Fe}{2} $\lambda5169$. We fail to fit the
excessive intensity of supposed \ion{He}{1}/\ion{Fe}{2} P~Cygni
feature around $5000{\rm~\AA}$. NLTE effects may account for it.
The excellent fit to the observed multi-peak shape between
$3800-4400{\rm~\AA}$ seems striking, considering numerous
transitions of \ion{Fe}{1}, \ion{Fe}{2}, \ion{Co}{2}, and
\ion{Ni}{2} in operation and complicated blending effects between
them. That is to say, our introduction of \ion{Ni}{2},
\ion{Co}{2}, and \ion{Fe}{1} is somewhat valid. Note the
\ion{Fe}{1} feature just redward to H$_\gamma$ in Figure
\ref{fig3} and other \ion{Fe}{1} features in Figure \ref{fig4}.

Overall, line identifications change little from day 37 to day
107, which means that there is no evident composition
stratification in the velocity range from $7,000{\rm~km~s^{-1}}$
to $3,000{\rm~km~s^{-1}}$. The spectra are fully dominated by
[\ion{O}{1}] and [\ion{Ca}{2}] nebular emissions since 1997 April
\citep{qiu99,mat01}, when the photosphere has most likely receded
into the oxygen core. Analysis of subsequent nebular spectra goes
beyond the ability of SYNOW.

\section{Discussion}

\subsection{Evolution of Photospheric Velocity}

The most reliable parameter fixed by the spectral synthesis with
SYNOW is the photospheric velocity. Absorption minima of some weak
lines, e.g. \ion{Fe}{2} $\lambda5018$, $\lambda5169$, and
H$_\gamma$, are often regarded as good tracers of $v_{\rm ph}$ in
Type II \citep[e.g.][]{eas89,dus95}. However, they more or less
blend with others and sometimes can not be identified correctly.
With SYNOW, information not only on minima positions but also on
profiles, and of as many as possible features, can be used to
determine $v_{\rm ph}$ in a consistent way, while line blending
and optical depth effects have been taken into account.

\begin{figure}
\plotone{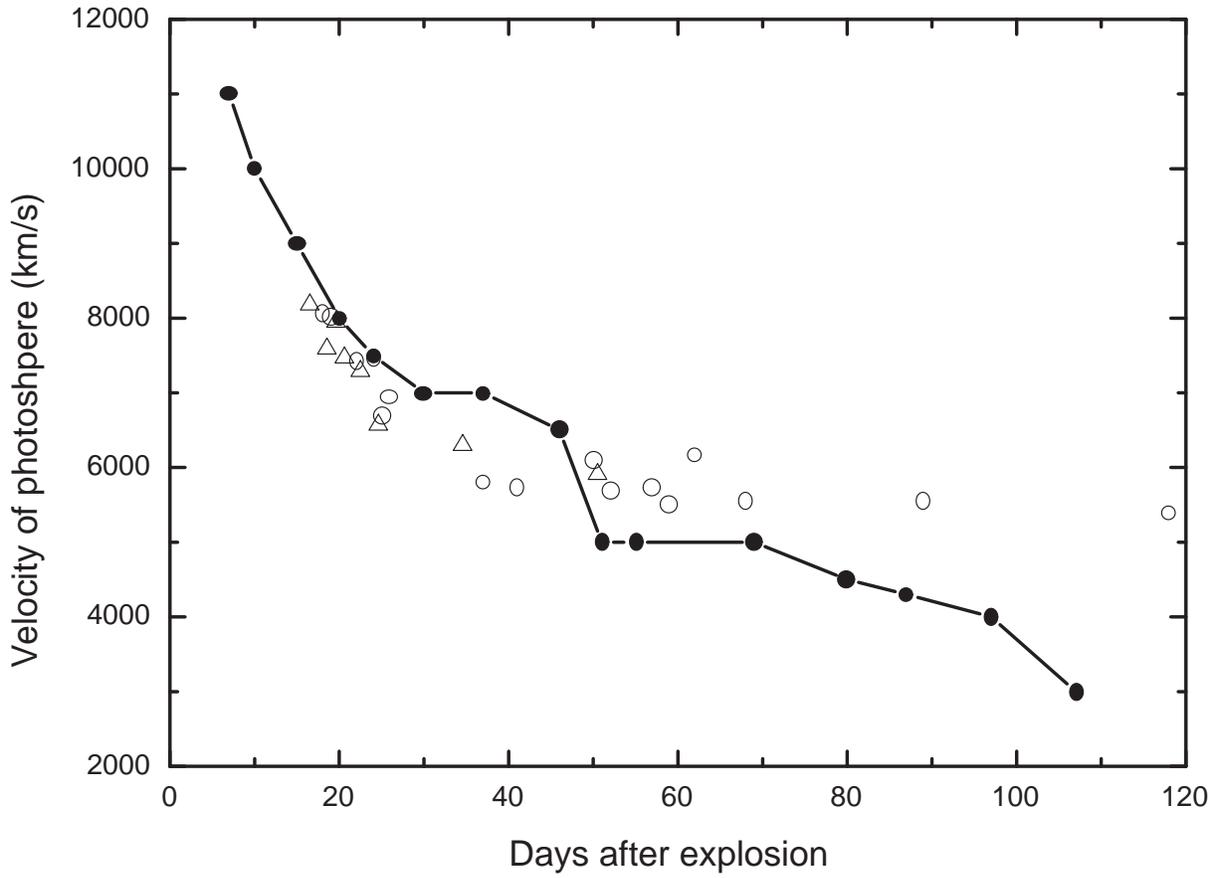} \caption{Photospheric velocities of SN 1996cb
(filled circles) compared with those of SN 1993J measured by
Prabhu et~al. (1995, open triangles) and Barbon et~al.(1995, open
circles). \label{fig5}}
\end{figure}

We plot in Figure \ref{fig5} the the evolution of photospheric
velocity of SN 1996cb, derived from our spectral synthesis. The
error is estimated less than $\pm 500{\rm~km~s^{-1}}$. We can see
that $v_{\rm ph}$ declines monotonically from $11,000
{\rm~km~s^{-1}}$ on day 7 to $3,000 {\rm~km~s^{-1}}$ on day 107
days. For comparison, photospheric velocities of SN 1993J measured
in absorption minimum of \ion{Fe}{2} $\lambda5018$
\citep{pra95,bar95b} are plotted in the same figure. They are
similar to each other on the whole, while the difference at late
phase may demonstrate that \ion{Fe}{2} $\lambda5018$ does not
follow $v_{\rm ph}$ very well (note that measurement of the same
line in our SN 1996cb spectra also gives a value
$\gtrsim6,000{\rm~km~s^{-1}}$ after day 80).

For given ejecta mass $M_{\rm ej}$ and explosion kinetic energy
$E$, we know $v_{\rm ph}$ roughly scales as $(E/M_{\rm
ej})^{1/2}$. The exact relationship depends on the specific
hydrodynamical structure of supernova ejecta. Considering the
exponential density profile adopted for our spectral synthesis
\begin{equation}
\rho=\rho_0\cdot\exp(-v/v_e)~,
\end{equation}
where $\rho_0$ is the central density and $v_e$ is the e-folding
velocity, one immediately gets
\begin{equation}
E=6M_{\rm ej}v_e^2~.
\end{equation}
Defining $\overline{\kappa}$ as the average opacity above the
photosphere, we find that the evolution of photospheric velocity
can be approximately expressed as
\begin{equation}
v_{\rm ph}\approx-v_e\ln\left[10^{-6}\frac{\tau_{\rm
ph}}{\overline{\kappa}} v_{e,3}^2(\frac{M_\sun}{M_{\rm
ej}})\right]-2v_e\ln t_{\rm d}~,
\end{equation}
where $\tau_{\rm ph}\approx 1$ or 2/3 is the optical depth at the
photosphere, $v_{e,3}$ is $v_e$ in $10^3{\rm~km~s^{-1}}$ and
$t_{\rm d}$ is time in days since explosion. Fitting of data shown
in Figure \ref{fig5} by the above equation gives
$v_e\approx1,400{\rm~km~s^{-1}}$, a little smaller than our trial
value of $2,000{\rm~km~s^{-1}}$ for spectral synthesis, and
\begin{equation}
M_{\rm ej}\approx4 M_{\sun}\cdot\tau_{\rm
ph}\cdot(\frac{\overline{\kappa}}{0.07})~,
\end{equation}
\begin{equation}
E\approx0.9\times10^{51}{\rm~ergs}\cdot(\frac{M_{\rm ej}
}{4~M_\sun})~.
\end{equation}

However, the actual density profile may be far from an exponential
one (In fact, $v_{\rm ph}(t)$ shown in Figure 5 can also be well
fitted by assuming a power-law density profile). Referring to the
4H89 model for SN 1993J presented by \citet[Fig. 2]{shi94}, in
which the density distribution approximates to $\rho\propto
v^{-6}$ between $24,000-5,500{\rm~km~s^{-1}}$ and remains roughly
constant below $5,500{\rm~km~s^{-1}}$, we then assume in SN 1996cb
\begin{equation}
\rho=\cases{\rho_c\cdot(\frac{v}{v_c})^{-n}~, & $v>v_c$~; \cr
\rho_c~, & $v\leqslant v_c$~.}
\end{equation}
Integrating $\rho$ and $\rho v^2$ over velocity, we get
\begin{equation}
E=\frac{3(n-3)}{5(n-5)}v_{c,4}^2(\frac{M_{\rm ej}}{M_\sun})\times
10^{51}{\rm~ergs}~,
\end{equation}
where $v_{c,4}$ is the characteristic velocity in
$10^4{\rm~km~s^{-1}}$. We find the evolution of photospheric
velocity can be described as
\begin{equation}
v_{\rm ph}=\cases{v_c\cdot(\frac{t}{t_c})^{-2/(n-1)}~, & $t>t_c$~;
\cr \frac{n\cdot v_c}{n-1}-\frac{v_c}{n-1}\cdot(\frac{t}{t_c})^2,
& $t\leqslant t_c$~,}
\end{equation}
which shows that $(t_c,v_c)$ is a inflection point. If we suppose
the bump from day 30 to 50 in Figure \ref{fig5} is more or less
relevant to the so-called inflection point, we then estimate $v_c$
at $\sim7,000-5,500{\rm~km~s^{-1}}$. By fitting $v_{\rm
ph}(t<t_c)$, we find $n\approx7.4$ for $v_c=7,000{\rm~km~s^{-1}}$
and $n\approx7.2$ for $v_c=5,500{\rm~km~s^{-1}}$. Substituting
these values into equation (8), one finally obtains
\begin{equation}
E\approx1-1.6\times10^{51}{\rm~ergs}\cdot(\frac{M_{\rm ej}
}{3~M_\sun})~.
\end{equation}

We know the typical value of explosion kinetic energy for
core-collapse supernova is $1\times10^{51}{\rm~ergs}$, accordingly
equations (9) and (5) suggest that ejecta mass of SN 1996cb
possibly lies in the range of $2-5~M_\sun$, similar to those of SN
1993J and Type Ib \citep{shi90}. More reasonable constraints on
$E$ and $M_{\rm ej}$ can be made by exploding a realistic
progenitor model and reproduce both light curves and $v_{\rm
ph}(t)$ shown in Figure \ref{fig5}.

\subsection{He-rich Hydrogen Envelope with High Velocity}

It is crucial to determine the distribution and mass of hydrogen
in Type IIb supernovae for the understanding of their unusual
pre-explosion evolution. For SN 1993J, most authors doing
hydrodynamical calculations inferred a low-mass He-rich hydrogen
envelope of $\sim0.1-1~M_\sun$
\citep[etc]{nom93,pod93,shi94,bar94b,utr94,woo94,you95}, which is
consistent with NLTE modeling of photospheric spectra
\citep{swa93b,zha96}. By analyzing its H$_\alpha$ emission in
nebular spectra, \citet{pat95} concluded that ionized hydrogen is
distributed between $7,500-11,400{\rm~km~s^{-1}}$ with density
maximum at $8,900-9,300{\rm~km~s^{-1}}$. \citet{utr96} arrived at
a very similar conclusion in his light curve study including
nonthermal ionization. Applying a sophisticated NLTE code to
nebular spectra, \citet{hou96} argued that the bulk of the
preferred H/He envelope mass, $0.2-0.4~M_\sun$, lies between
$8,500-10,000{\rm~km~s^{-1}}$, and ruled out the presence of more
than $0.02~M_\sun$ of hydrogen above or below this range.

As described above, to fit photospheric spectra of SN 1996cb, the
line forming region of hydrogen is required to be detached from
the photosphere after day 10. The adopted value of parameter
$v_{\rm nin}^{\rm HI}$ only varies in a very narrow range,
$9,500-10,500{\rm~km~s^{-1}}$. It is about $1,000{\rm~km~s^{-1}}$
smaller than the absorption velocity of H$_\alpha$ till day 46;
but after that it approaches the latter. In comparison, absorption
velocity of H$_\alpha$ and H$_\beta$ in SN 1993J, considerably
declining during the first 50 days, stop around
$9,500{\rm~km~s^{-1}}$ and $8,500{\rm~km~s^{-1}}$ respectively
from day 50 to day 70 \citep{bar95a}. After day 100, as H$_\alpha$
absorption is no longer optical thick, the velocity reaches
$8,500{\rm~km~s^{-1}}$, i.e. the minimum velocity of hydrogen
envelope derived by \citet{hou96}. Accordingly, we can say that
the inner boundary of hydrogen envelope of SN 1996cb exists at
$9,500-10,500{\rm~km~s^{-1}}$, higher than of SN 1993J.

In the explosion of a Type IIb supernova, due to the supposed low
mass of hydrogen envelope, the deceleration of shock wave at the
H/He interface and consequent Rayleigh-Taylor instability is
rather weak \citep{iwa97}. As a result, the minimum velocity of
hydrogen envelope depends mainly on its mass, $M_{\rm env}$. Thus
one can expect an even smaller $M_{\rm env}$ for SN 1996cb than
for SN 1993J. Considering an ejecta with the exponential density
profile given by equation (1), the mass above $v_{\rm min}^{\rm H}
\approx9,500-10,500{\rm~km~s^{-1}}$, i.e. $M_{\rm env}$, is
\begin{eqnarray}
M_{\rm env} & = & \left[\frac{1}{2}f^2+f+1\right]e^{-f}\cdot
M_{\rm ej}  \\
& \approx & 0.02-0.04~M_{\rm ej}~,
\end{eqnarray}
where $f\equiv v_{\rm min}^{\rm H}/v_e$. And for the density
profile given by equation (7), $M_{\rm env}$ is
\begin{eqnarray}
& & M_{\rm env} = \frac{3}{n}\cdot(\frac{v_c}{v_{\rm min}^{\rm H}
})^{n-3} \cdot
M_{\rm ej} \\
& \approx & \cases{0.07-0.1~M_{\rm ej}~, & $v_{c,4}=0.7$~; \cr
0.03-0.04~M_{\rm ej}~, & $v_{c,4}=0.55$~.}
\end{eqnarray}
Combining equations (11) and (13) with equations (5) and (9)
respectively, and assuming $E\approx1\times10^{51}{\rm~ergs}$, we
find the mass of hydrogen envelope of SN 1996cb is $M_{\rm
env}\sim0.1-0.2~M_\sun$, near the low end among different
estimated values for SN 1993J.

As argued above, the photosphere in SN 1996cb recedes through the
H/He interface around day 10, much earlier than in SN 1993J, say
day 26. This could be explained by the smallness of $M_{\rm env}$
compared with SN 1993J. It is well known that, opacity in the
hydrogen envelope is dominated by free electron scattering at the
early phase. Then the optical depth at the the H/He interface at
time $t$ after explosion roughly scales as
\begin{equation}
\tau\sim n_e\sigma_T\ell~,
\end{equation}
where $\ell\sim v_{\rm min}^{\rm H}t$ is the characteristic length
in the hydrogen envelope, $n_e$ is the characteristic electron
density and $\sigma_T$ the Thompson scattering cross section. It
is nonthermal ionization by relativistic electrons that prevents
the hydrogen envelope from being completely recombined after the
initial rapid cooling stage \citep{whe96,utr96}. Adopting a simple
balance between collisional recombination and nonthermal
ionization,
\begin{equation}
\alpha n_e N^+\sim\Gamma N~,
\end{equation}
and neglecting any variation among the effective collisional
recombination coefficient $\alpha$ and the effective nonthermal
ionization coefficient $\Gamma$, one gets
\begin{equation}
n_e\sim N^+\propto\sqrt{N}\sim\sqrt{\frac{M_{\rm env}}{\ell^3}}~.
\end{equation}
When the photosphere arrives at the H/He interface, i.e.
$\tau\approx 1$, substituting of equation (16) into equation (14)
gives
\begin{equation}
t\propto\frac{M_{\rm env}}{v_{\rm min}^{\rm H}}~.
\end{equation}
It is obvious that this equation is in agreement with the
comparison between SN 1996cb and SN 1993J.

Helium abundance $Y$ in the so-called hydrogen envelope should be
very high, which is suggested by strong \ion{He}{1} $\lambda5876$
feature in the spectrum of SN 1996cb on day 7, when the
photosphere still stays in the envelope. We know the line Sobolev
optical depth is
\begin{equation}
\tau\approx0.23f\cdot\lambda(\micron)\cdot N_{\ell}({\rm
cm^{-3}})\cdot t_{\rm d}~,
\end{equation}
where $f$ is the oscillator strength, $\lambda$ is the rest
wavelength, $N_{\ell}$ is the atomic number density in the lower
state, $t_{\rm d}$ is time in days after explosion and the term
for induced emission is neglected. Under the LTE assumption, the
adopted optical depths of reference lines and excitation
temperatures for \ion{H}{1} and \ion{He}{1} give $N_{\rm
HI}\sim10^6{\rm~cm^{-3}}$ and $N_{\rm HeI}\sim10^9{\rm~cm^{-3}}$,
respectively, at the photosphere on day 7, while the electron
density required for continuum optical depth is
$n_e\sim10^{10}{\rm~cm^{-3}}$. These values, although rather
uncertain, may show both that the helium abundance is high and a
large fraction of H and He are ionized. Through NLTE modeling of
early spectra, \citet {bar94a} found $Y\approx0.8$ for the
envelope of SN 1993J.

A He-rich hydrogen envelope is the natural result of pre-supernova
evolution with large mass loss \citep{sai88,woo94}. For given
$M_{\rm core}$, $M_{\rm env}$, $Y$ and surface radius $R_0$, a
hydrostatic and thermal equilibrium H/He envelope can be
constructed upon the helium core \citep{sai88}. Since the initial
adiabatic cooling stage in SN 1996cb ends before the discovery,
i.e. day $3$, much earlier than in SN 1993J, i.e. day $9$, $R_0$
of the progenitor of SN 1996cb should by smaller than that of SN
1993J \citep[$\sim400~R_\sun$,][]{iwa97}. A solution of $(M_{\rm
core}; M_{\rm env},Y,R_0)$ that can fit both the light curve and
the day 7 spectrum will discriminate the pre-explosion paths of SN
1996cb and SN 1993J.

We note before the excessive emission of H$_\alpha$ in day 7
spectrum and its blueshifted peak can not be fitted with SYNOW. As
in the atmospheres of normal Type II, it can be attributed to NLTE
effects and the significant extension of the line forming region
\citep[e.g.][]{dus95,eas96}. Collisional recombination of H$^+$ to
$n\geqslant3$ levels and subsequent cascade transitions may also
contribute. \citet{zha95} introduced a smooth but tenuous outer
layer of the envelope to model the unusually broad H$_\alpha$
absorption profile in early spectra of SN 1993J, which also
appears in day 7 spectrum of SN 1996cb but is less prominent and
hence suggests a somewhat different density profile for the outer
layer.

\subsection{Identifications of \ion{Ni}{2} and \ion{Co}{2} Lines}

We find that \ion{Ni}{2} significantly contributes to an
absorption feature near $\sim4000{\rm~\AA}$, i.e. adjacent to
\ion{Ca}{2} H\&K. These lines, mostly \ion{Ni}{2} $\lambda\lambda
4016,4067$, have been identified unambiguously in photospheric
spectra of some Type Ia \citep[e.g.][]{maz97,hat99b}. This is not
unexpected since in thermonuclear explosions of white dwarfs Ni is
synthesized in large abundance and distributed outward to rather
high velocity. As to core-collapse supernovae, \citet{den00}
introduced \ion{Ni}{2} to account for the ``$3930{\rm~\AA}$
absorption'' in their September 14 spectrum of Type Ib SN 1999dn,
while \citet{maz00} found that \ion{Ni}{2} and \ion{Co}{2} lines
help to form the unusual spectra of Type Ic hypernova SN 1997ef.

Ni identified in SN 1996cb is probably primordially originated.
For core-collapse supernovae, typically $\sim 0.1{\rm~M_\sun}$
newborn $^{56}$Ni is ejected but, if without mixing, buried in the
innermost layer of the ejecta. On the other hand, artificial
large-scale outward mixing of $^{56}$Ni plays an important role in
reproducing observed light curves \citep[e.g.][]{bli00} and
spectra \citep[e.g.][]{luc91} with one-dimensional models. 2D and
3D hydrodynamical numerical simulations demonstrated that
large-scale mixing can be induced by nonlinear growth of
Rayleigh-Taylor instabilities and can transport $^{56}$Ni to
$\lesssim2000{\rm~km~s^{-1}}$ for Type II SNe
\citep[e.g.][]{kif00} and to $\lesssim7000{\rm~km~s^{-1}}$ for
Type Ib/IIb SNe \citep[e.g.][]{hac91,iwa97}. Accordingly,
\ion{Ni}{2} $4000{\rm~\AA}$ feature identified in SN 1996cb as
early as day 7, when the photosphere is well in the hydrogen
envelope and with a velocity as high as $11,000{\rm~km~s^{-1}}$,
is unlikely to be produced by newborn $^{56}$Ni. Additional
evidence comes from the fact that the \ion{Ni}{2} feature remains
discernible far beyond the half life of $^{56}$Ni, $\sim6$ days.

An alternative explanation for high velocity \ion{Ni}{2} is
asymmetry. \citet{nag98} and \citet{mae00} have calculated
nucleosynthesis for axisymmetric supernova explosions and
concluded that $^{56}$Ni can be accelerated to very high
velocities roughly along the axial direction. A bipolar scenario
of explosion for Type IIb supernova is also the suggestion of
\citet{wan01}, who noticed the strikingly similarity of
spectropolarimetry between SN 1993J and SN 1996cb.

The identification of \ion{Co}{2} in SN 1996cb seems more reliable
than that of \ion{Ni}{2}. Distinct \ion{Co}{2} features emerged in
the blue part of spectra since day 37, when the photosphere
velocity declined to $7,000{\rm~km~s^{-1}}$. This can be regarded
as direct spectral evidence of large-scale outward mixing of
radioactive elements in explosion. We mentioned earlier that
similar but less distinct \ion{Co}{2} features may occur in SN
1993J after day 56, when the photospheric velocity is lower than
$6,000{\rm~km~s^{-1}}$. This is consistent with the results of
light curve modeling \citep{iwa97} and NLTE analysis of nebular
spectra \citep{hou96}, that maximum velocity of $^{56}$Ni for SN
1993J most likely lies in the range of
$\sim4,000-6,000{\rm~km~s^{-1}}$. As shown by \citet{hac91} and
\citet{iwa97}, the extent of $^{56}$Ni mixing provides a
constraint on the mass of helium core of the progenitor,
$M_\alpha$, that more extensive mixing is induced by R-T
instabilities for smaller $M_\alpha$, and that $v_{\max}^{\rm
Ni}\sim6,000{\rm~km~s^{-1}}$ for $M_\alpha=3.3~M_\sun$ and
$v_{\max}^{\rm Ni}\sim3,000{\rm~km~s^{-1}}$ for
$M_\alpha=4~M_\sun$. Accordingly, the mass of helium core of SN
1996cb may be $\lesssim 3.3~M_\sun$, i.e. a little smaller than
that of SN 1993J. The more extensive mixing of $^{56}$Ni also
helps to explain the earlier start of the radioactive heating
stage in the light curve of SN 1996cb than in that of SN 1993J.

\subsection{Blueshift of [\ion{O}{1}] $\lambda 5577$ Emission Peak}

We identify the distinct $\sim5500{\rm~\AA}$ bump, in photospheric
spectra of SN 1996cb from day 30 onwards, as blueshifted
[\ion{O}{1}] $\lambda5577$ emission. The appearance of this line
at such early phase was first mentioned by \citet{har87} in Type
Ib SNe 1983N and 1984L. But \citet{fil90} argued that spectral
synthesis is required to make a reliable identification, when
analyzing spectra of Type Ic SN 1987M \citep[see also][]{swa93a}.
\citet{wan94} found that in SN 1993J this feature can be traced
back to as early as day 27 and also noticed its apparent
blueshift, which they attributed to a clumpy nature of ejecta near
the He/C+O interface. \citet{fil93} mentioned that the blueshift
of emission lines is just a consequence of viewing primarily the
near side of the optically thick ejecta. \citet{spy94} compared
peak wavelengths of [\ion{O}{1}] lines and \ion{Mg}{1}] 4571 with
those of permitted lines and proposed a large-scale asymmetric
distribution of radioactive material for SN 1993J. However,
through modeling nebular spectra of SN 1993J, \citet{hou96}
concluded that the bulk of the blueshift of [\ion{O}{1}] lines can
be explained by line blending effects. Blueshift of the order of
$1,000{\rm~km~s^{-1}}$ is also observed for emission lines at late
phases of Type Ib SN 1996N \citep{sol98}.

\begin{figure}
\plotone{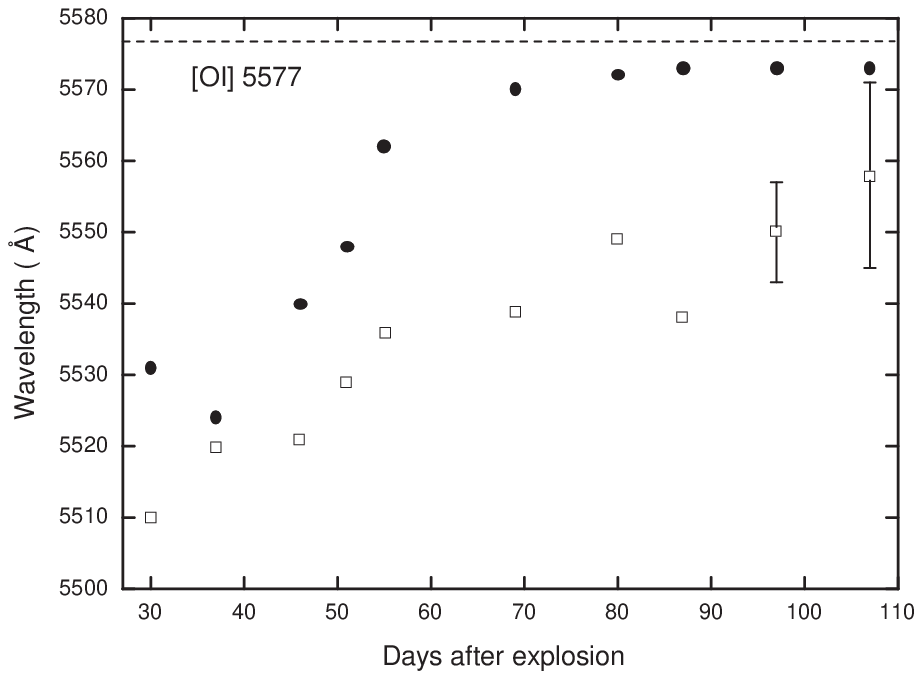} \caption{Peak wavelength of [\ion{O}{1}]
$\lambda5577$ in observed spectra (open squares) compared with
that in best fit synthetic spectra (filled circles). \label{fig6}}
\end{figure}

To study the effect of line blending on the blueshift of
[\ion{O}{1}] $\lambda5577$ at photospheric phases, we try to
include [\ion{O}{1}] $\lambda5577$ and $\lambda\lambda6300,6364$
into our synthetic spectra by setting an excitation temperature as
low as $4,500{\rm~K}$. This technique is roughly valid because the
emission peak of a pure resonant scattering line, in a spherically
symmetric and homologous expanding atmosphere, exists at the rest
wavelength if without blending \citep{jef90}, although in fact
[\ion{O}{1}] net emissions is mainly produced via other processes,
like electron collisional excitation. Peak wavelength of
[\ion{O}{1}] $\lambda5577$ both in observed spectra and in our
best fit synthetic ones are plotted in Figure \ref{fig6}. These
values are simply measured by hand, for the accuracy is not
crucial here. On day 97 and 107, some features or noise develop on
the top of [\ion{O}{1}] $\lambda5577$, so we introduce error bars
to describe the possible large uncertainty.

Figure \ref{fig6} shows that the observed blueshift declines
steadily from $\sim70{\rm~\AA}$ on day 30 to $\sim20{\rm~\AA}$ on
day 107. This trend is reproduced in the rough by our synthetic
data, which can only be interpreted as the variation of blending
effects because neither asymmetry nor clumping is involved in our
spectral synthesis. On the other hand, a blueshift of
$\sim20{\rm~\AA}$ of the observed peak wavelength to the synthetic
one remains throughout the period covered by Figure \ref{fig6},
while the latter approaches to $5577{\rm~\AA}$ finally.
Furthermore, if we check the nebular spectra in April or May of
1997 \citep{qiu99,mat01}, we can find that [\ion{O}{1}]
$\lambda5577$ is still blueshifted by $\sim20{\rm~\AA}$.

The bulk of the blueshift of [\ion{O}{1}] $\lambda5577{\rm~\AA}$
at photospheric phases in SN 1996cb can be accounted for by line
blending, but the rest of $\sim20{\rm~\AA}$, i.e.
$\sim1000{\rm~km~s^{-1}}$, still demands an explanation. (1) The
clumping model of \citet{wan94} seems less likely because the
[\ion{O}{1}] emission peak never return to the rest wavelength,
even after the photosphere has receded deep into the oxygen core.
However, a photospheric spectrum code which can manage a clumpy
ejecta well, e.g. by using the Monte Carlo technique, is required
to clarify if clumping can affect the peak wavelength of line
emission more or less. (2) At photospheric phases, the occultation
of the main receding part of emission line forming region by the
photosphere will also contribute somewhat to the residual
blueshift, especially if NLTE effects can remarkably populate the
upper level of [\ion{O}{1}] $\lambda5577$ transition, $\rm
2p^4(^1S)$, well above the photosphere. (3) \citet{hou96} and
\citet{sol98} found the blueshift of \ion{O}{1} $\lambda777$ in
nebular spectra of SN 1993J, $\lesssim 500{\rm~km~s^{-1}}$, and of
SN 1996N, $\sim1000{\rm~km~s^{-1}}$, respectively. Since this line
is probably free of strong blending, they suggested that it is
indicative of real asymmetry, possibly related to large-scale
mixing.

In photospheric spectra of SN 1996cb, \ion{O}{1} $\lambda7773$ is
inconspicuous and contaminated by telluric absorption.
[\ion{O}{1}] $\lambda\lambda6300,6364$ photons are strongly
blended with each other and scattered by H$_\alpha$ transition.
Therefore the measurement of peak wavelengths of these lines is of
no significance. On the other hand, the distinct \ion{Mg}{1}]
$\lambda4571$ narrow emission in our day 51 and 87 spectra peaks
at $\sim4592{\rm~\AA}$, i.e. apparently redshifted. Does features
of \ion{Mg}{1}] and [\ion{O}{1}] come from different asymmetry and
hence hint different mixing behavior? The peak wavelength of
prominent [\ion{Ca}{2}] $\lambda\lambda7291,7324$ emission evolves
from $7304{\rm~\AA}$ on day 51 to $7324{\rm~\AA}$ on day 107.
Considering the blending with \ion{He}{1} $\lambda7281$ in the
beginning, it is just to say that no blueshift exists \citep[cf.
SN 1996N][]{sol98}. This is consistent with the conclusion for SN
1987A that calcium is most likely primordial and much less clumpy
than oxygen \citep{li92,li93}.

Is the electron density in SN 1996cb as early as day 30 low enough
to favor the emission of oxygen forbidden lines? We know the
critical electron density for [\ion{O}{1}] $\lambda5577$ forbidden
transition is
\begin{equation}
N_e^{\rm crit}=\frac{g_u
A_{u\ell}}{8.6\times10^{-6}T_e^{-1/2}\Omega_{u\ell}(T_e)}~,
\end{equation}
where the statistical weight $g_u=5$, the spontaneous radiative
transition rate $A_{u\ell}\approx1.22{\rm~s^{-1}}$, $T_e$ is the
electron temperature, and the effective collision strengths
$\Omega_{u\ell}$ is found from \citet{bha95}. The calculated
$N_e^{\rm crit}$ is insensitive to $T_e$ and, when
$T_e\sim5,000{\rm~K}$, $N_e^{\rm
crit}\sim7\times10^8{\rm~cm^{-3}}$. Assuming pure Thompson
scattering continuum optical depth, we can express the electron
density at the photosphere as
\begin{equation}
n_e^{\rm ph}=\frac{\tau_{\rm ph}}{\sigma_e v_e
t}\approx\frac{2\times10^{11}{\rm~cm^{-3}}}{v_{e,3} t_{\rm d}}
\end{equation}
for the density profile defined by equation (1), and as
\begin{equation}
n_e^{\rm ph}=\frac{(n-1)\tau_{\rm ph}}{\sigma_e v_{\rm ph}
t}\approx\frac{(n-1)\cdot2\times10^{11}{\rm~cm^{-3}}}{v_{{\rm
ph},3} t_{\rm d}}
\end{equation}
for the density profile defined by equation (6). Both equations
give $N_e^{\rm ph}\sim5\times10^9{\rm~cm^{-3}}$ on day 30, which
does not differ much from $N_e^{\rm crit}$ for [\ion{O}{1}]
$\lambda5577$. We note that \citet{swa93b} exploit a low-density
outer layer in high velocity, say $\>12,000{\rm~km~s^{-1}}$, to
fit oxygen features in photospheric spectra of SN 1993J. This
assumption seems both unnecessary and unreasonable.

\begin{figure}
\plotone{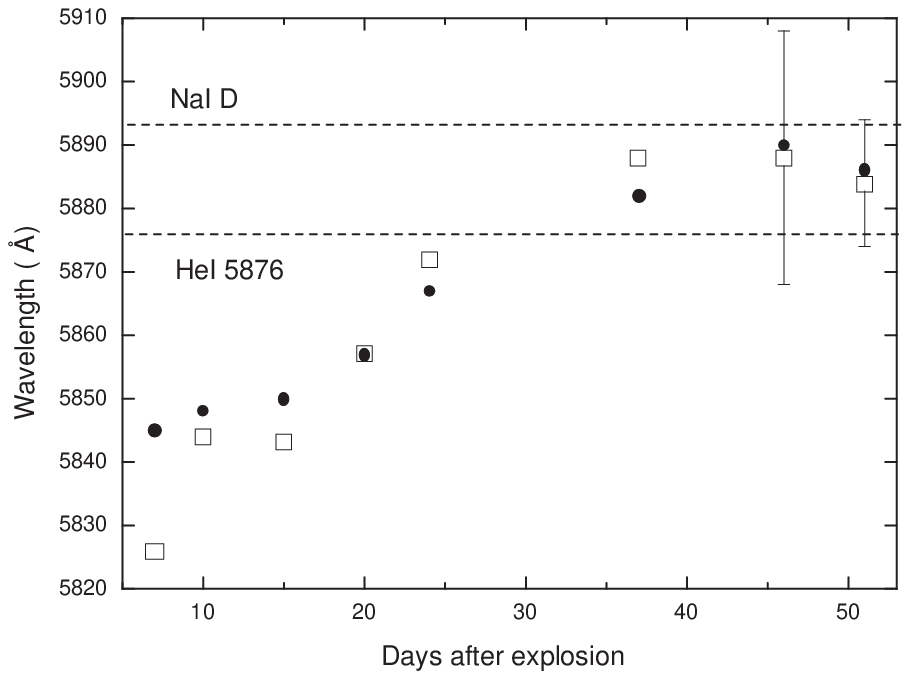} \caption{Peak wavelength of \ion{He}{1}
$\lambda5876$ in observed spectra (open squares) compared with
that in best fit synthetic spectra (filled circles). \label{fig7}}
\end{figure}

\citet{qiu99} also noticed the apparent blueshift of \ion{He}{1}
$\lambda5876$ emission peak at very early phases and its
subsequent recession to the rest wavelength in SN 1996cb. They
claimed it as evidence of prominent Rayleigh-Taylor instabilities
at the H/He interface. However, according to two-dimensional
simulations, R-T instabilities must be weak at the H/He interface
of Type IIb because of small envelope mass \citep[e.g.][]{iwa97}.
We plot in Figure \ref{fig7} peak wavelength of \ion{He}{1}
$\lambda5876$ both in observed spectra and in our best fit
synthetic ones. The observed values can be reproduced very well by
our synthetic spectra. Obviously, the blueshift of \ion{He}{1}
$\lambda5876$ emission peak is superficial and can be attributed
to the strong blending effects with H$_{\alpha}$. After day 37,
\ion{He}{1} $\lambda5876$ emission peak is redshifted into the
range of $5876-5893{\rm~\AA}$, which shows that contribution from
\ion{Na}{1} D becomes noticeable.

\subsection{Some Ambiguous Identifications}

Three distinct notches at $4290{\rm~\AA}$, $4370{\rm~\AA}$, and
$4610{\rm~\AA}$ appears on day 30 spectrum and have been
identified as \ion{V}{1} absorption, \ion{Ba}{2} $\lambda4554$,
and \ion{C}{2} $\lambda\lambda4738,4744$, respectively. With some
caution, they can be traced back to day 24 spectrum. However, we
must note here that these identifications are rather ambiguous.

First, the ionization potential of \ion{V}{1}, only $6.7{\rm~eV}$,
seems too small to prevent it from being largely ionized.
Therefore the $4290{\rm~\AA}$ feature is more likely attributed to
\ion{Ti}{2} absorption, just like in SN 1987A \citep{jef90}, Type
Ic SN 1994I \citep{mil99}, and Type Ic hypernova SN 1997ef
\citep{maz00}, etc. However, with \ion{Ti}{2} we can not produce
the required notch but lower the flat top of H$_\gamma$ uniformly,
unless we set a strong constraint on the line forming region, say
$v_{\rm max}^{\rm TiII}\lesssim 10,000{\rm~km~s^{-1}}$. This does
not necessarily rule out this ion, because many \ion{Ti}{2} lines
in strongly blending is involved and NLTE effects if existed will
complicate the case greatly.

At first sight, the identification of \ion{Ba}{2} $\lambda4554$ is
more or less credible, since the observed feature is prominent and
another absorption feature can be fitted by \ion{Ba}{2}
$\lambda6142$ consistently. But the required minimum velocity of
the line forming region, $v_{\rm min}^{\rm
BaII}=13,000{\rm~km~s^{-1}}$, is far from explicable. How can the
barium material obtain significant resonant scattering optical
depth when the photosphere is $6000{\rm~km~s^{-1}}$ below,
considering that both the ejecta and the radiation field have been
greatly diluted compared with the photosphere? Is it produced by a
mass of dense barium cloudy which run into our line of sight from
day 24 to day 37, say by the rotation of supernova? But the
required rotational velocity is $>10^3{\rm~km~s^{-1}}$ and hence
impossible. \citet{mat00a} suggested that a line at
$\sim4430{\rm~\AA}$ on day 19 spectrum of SN 1993J may be
\ion{Ba}{2} $\lambda4554$. But according to our spectral synthesis
for SN 1996cb, it is undoubtedly \ion{Fe}{2} feature (see day 7
spectrum shown in Figure \ref{fig1}).

To fit the $4610{\rm~\AA}$ dip in the spectra of day 24 and 30 by
\ion{C}{2} $\lambda\lambda4738,4744$, we make \ion{C}{2} to be
detached at $v_{\rm min}=9,800{\rm~km~s^{-1}}$. However, from day
37 onwards, the feature formed in this way is too blue to fit the
observed one. Inspecting these spectra closely, we find that this
dip always attaches to the blueward edge of H$_\beta$ absorption
that it looks like a satellite line. We note that \citet{bar00}
met the same problem of the lack of a strong candidate for the
$4610{\rm~\AA}$ feature when analyzing the spectra of Type II SN
1999em. By using a full NLTE code PHOENIX, they found that it is
produced by complicated NLTE effects varying the Balmer level
populations in the mid-velocity range. They named it and H$_\beta$
together ``double H$_\beta$''. It is interesting to note that
\citet{utr95} got a close idea, namely a radial dependence of
Sobolev optical depth with two maxima, to explain the blue
emission satellite of the famous ``Bochum event'' in SN 1987A. The
recurrence of this phenomenon in these three otherwise quite
different supernova arouses the question how generally it would
exist.

Since SN 1987A \citep{wil87}, \ion{Sc}{2} and \ion{Sr}{2} features
are identified in many other Type II supernovae, e.g. SN 1992H
\citep{clo96}, SN 1995V \citep{fas98}, and SN 1997D \citep{tur98},
and often coexist with \ion{Ba}{2} ones. Can the
$\sim5500{\rm~\AA}$ bump here be identified as \ion{Sc}{2}
$\lambda5527$, especially in the day 30 spectrum? We invoke
\ion{Sc}{2} in our synthesis calculations but find that the
synthetic emission feature peaks at $5490{\rm~\AA}$, $20{\rm~\AA}$
redder than the observed one on day 30, i.e. $5510{\rm~\AA}$.
Furthermore, the \ion{Sc}{2} $\lambda4247$ is much too strong if
the excitation temperature $\leqslant12,000{\rm~K}$. As shown in
Figure 1 (dotted line), \ion{Sr}{2} $\lambda4078$ blending with
H$_\delta$ can fit the $\sim4000{\rm~\AA}$ notch the same well as
\ion{Ni}{2} lines on day 10, but in the price of an unwanted
absorption feature due to \ion{Sr}{2} $\lambda4215$.

\section{Conclusion}

We have analyzed the photospheric spectra of Type IIb SN 1996cb
from day 7 to day 107 and made detailed line identifications by
using the parameterized supernova synthetic-spectrum code SYNOW.
Our findings are the following:
\begin{enumerate}
\item The photospheric velocity evolves from $11,000{\rm~km~s^{-1}}$ on
day 7 to $3,000{\rm~km~s^{-1}}$ on day 107, which makes some
constraint on the explosion kinetic energy and ejecta mass, i.e. $
E\approx0.7-1.6\times10^{51}{\rm~ergs}\cdot(M_{\rm ej}/3M_\sun)$.
\item The minimum velocity of hydrogen envelope is $v_{\min}^{\rm
HI}\sim9,500-10,500{\rm~km~s^{-1}}$, corresponding to a mass of
hydrogen envelope of $0.1-0.2~M_\sun$, smaller than that of SN
1993J.
\item Distinct \ion{Ni}{2} and \ion{Co}{2} features have been
identified. The former is most likely primordially originated. The
latter shows that newly synthesized radioactive elements has been
mixed outward to at least $7,000{\rm~km~s^{-1}}$, which favors a
mass of helium core of the progenitor of $\lesssim 3.3~M_\sun$.
\item The bulk of the blueshift of [\ion{O}{1}] $\lambda 5577$ net
emission is attributed to line blending, although a still
considerable residual $\sim20{\rm~\AA}$ remains till the late
phase. On the other hand, the superficial blueshift of \ion{He}{1}
$\lambda 5876$ peak can be fully explained as the blending effect
with H$_{\alpha}$.
\end{enumerate}

Although the results presented in this paper are calculated with a
parameterized code under the purely resonant scattering
assumption, they may serve as initial references for detailed NLTE
spectrum modeling, which is self-consistent and more reliable but
on the other hand time-consuming. The modeling of the light curve
of SN 1996cb, coupled with what have been derived from our direct
spectral analysis, like the evolution of photospheric velocity,
minimum velocity of hydrogen envelope and extent of $^{56}$Ni
mixing, is in progress.

\acknowledgements

We specially thank David Branch for providing us with the code
SYNOW, and for helpful discussions. We are grateful to Ken'ichi
Nomoto, Takayoshi Nakamura, Hideyuki Umeda, and Kazuhito Hatano
for discussions. Jinsong Deng is currently supported by the JSPS
Postdoctoral Fellowship for Foreign Researchers. This work has
been supported by Grants-in-Aid for Scientific Research from JSPS
and COE research (07CE2002) of the Japanese Ministry of Education,
Science, Culture and Sports, and by Chinese Natural Science
Foundation.

\clearpage

\clearpage

\clearpage

\clearpage

\clearpage

\clearpage

\clearpage

\clearpage

\begin{deluxetable}{lccr}
\tabletypesize{\scriptsize} \tablecaption{Epochs and fitting
parameters of the observed spectra \label{tbl1}} \tablewidth{0pt}
\tablehead{ \colhead{Date(UT)} & \colhead{Epoch} &
\colhead{$v_{\rm ph}~({\rm km~s^{-1}})$} & \colhead{$T_{\rm
bb}~({\rm~K})$} }
\startdata
1996 Dec 19 & 7/-14 & 11,000 & 10,500\\
1996 Dec 22 & 10/-11 & 10,000 & 9,700\\
1996 Dec 27 & 15/-6 & 9,000 & 9,300\\
1997 Jan 1 & 20/-1 & 8,000 & 10,500\\
1997 Jan 5 & 24/+3 & 7,500 & 7,500\\
1997 Jan 11 & 30/+9 & 7,000 & 6,200\\
1997 Jan 18 & 37/+16 & 7,000 & 5,200\\
1997 Jan 27 & 46/+25 & 6,500 & 5,500\\
1997 Feb 1 & 51/+30 & 5,000 & 5,000\\
1997 Feb 5 & 55/+35 & 5,000 & 5,000\\
1997 Feb 19 & 69/+48 & 5,000 & 5,000\\
1997 Mar 1 & 80/+59 & 4,500 & 6,000\\
1997 Mar 8 & 87/+66 & 4,300 & 6,000\\
1997 Mar 18 & 97/+76 & 4,000 & 5,500\\
1997 Mar 28 & 107/+86 & 3,000 & 6,400\\
\enddata
\end{deluxetable}


\begin{thebibliography}{}

\bibitem[Baron, Hauschildt, \& Branch(1994)]{bar94a} Baron,~E.,
    Hauschildt,~P.~H., \& Branch,~D. 1994, \apj, 426, 334
\bibitem[Baron et~al.(1995)]{bar95b} Baron,~E., Hauschildt,~P.~H.,
    Branch,~D., et~al. 1995, \apj, 441, 170
\bibitem[Baron et~al.(2000)]{bar00} Baron,~E., Branch,~D.,
    Hauschildt,~.P.~H., et~al. 2000, \apj, 545, 444
\bibitem[Barbon et~al.(1995)]{bar95a} Barbon,~R., Benetti,~S.,
    Cappellaro,~E., et~al. 1995, \aaps, 110, 513
\bibitem[Bartunov et~al.(1994)]{bar94b} Bartunov,~O.~S.,
    Blinnikov,~S.~I., Pavlyuk,~N.~N., et~al. 1994, \aap, 281, L53
\bibitem[Benetti et~al.(2000)]{ben00} Benetti,~S., Cappellaro,~E.,
    Turatto,~M., et~al. 2000, \iaucirc, 7375
\bibitem[Bhatia \& Kastner(1995)]{bha95} Bhatia,~A.~K., \&
    Kastner,~S.~O. 1995, \apj, 96, 325
\bibitem[Blinnikov et~al.(1998)]{bli98} Blinnikov,~S.~I.,
    Eastman,~R., Bartunov,~O.~S., et~al. 1998, \apj, 496, 454
\bibitem[Blinnikov et~al.(2000)]{bli00} Blinnikov,~S.~I.,
    Lundqvist,~P., Bartunov,~O., et~al. 2000, 532, 1132
\bibitem[Branch(1980)]{bra80} Branch,~D. 1980, in Proc. Workshop on Atom. Phys.
    and Spectrsc., Supernovae Spectra, ed. R.~Meyerott \& G.~H.~Gillespie (New York:
    AIP), 39
\bibitem[Chornock, Modjaz, \& Filippenko(2001)]{cho01} Chornock,~R.,
    Modjaz,~M., \& Filippenko,~A.~V. 2001, \iaucirc, 7618
\bibitem[Clocchiatti et~al.(1996)]{clo96} Clocchiatti,~A.,
    Benetti,~S., Wheeler,~J.~C., et~al. 1996, \aj, 111, 1286
\bibitem[Deng et al.(2000)]{den00} Deng,~J.~S., Qiu,~Y.~L., Hu,~J.~Y., et~al.
    2000, \apj, 540, 452
\bibitem[Duschinger et~al.(1995)]{dus95} Duschinger,~M., Puls,~J.,
    Branch,~D., et~al. 1995, \aap, 297, 802
\bibitem[Eastman \& Kirshner(1989)]{eas89} Eastman,~R.~G., \&
    Kirshner,~R.~P., 1989, 347, 771
\bibitem[Eastman, Schmidt, \& Kirshner(1996)]{eas96} Eastman,~R.~G.,
    Schmidt,~B.~P., \& Kirshner,~R. 1996, \apj, 466, 911
\bibitem[Fassia et~al.(1998)]{fas98} Fassia,~A., Meikle,~W.~P.~S.,
    Geballe,~T.~R., et~al. 1998, \mnras, 299, 150
\bibitem[Filippenko(1988)]{fil88} Filippenko,~A.~V. 1988, \aj, 96, 1941
\bibitem[Filippenko \& Shields(1989)]{fil89} Filippenko,~A.~V.,
    \& Shields,~J.~C. 1989, \iaucirc, 4851
\bibitem[Filippenko, Porter, \& Sargent(1990)]{fil90} Filippenko,~A.~V.,
    Porter,~A.~C., \& Sargent,~W.~L.~W. 1990, \aj, 100, 1575
\bibitem[Filippenko, Matheson, \& Barth(1993)]{fil93} Filippenko,~A.~V.,
    Matheson~T., Barth,~A.~J. 1994, \apj, 108, 2220
\bibitem[Filippenko(1997)]{fil97} Filippenko,~A.~V. 1997, \araa, 35, 309
\bibitem[Filippenko \& Chornock(2001)]{fil01} Filippenko,~A.~V., \& Chornock,~R. 2001,
    \iaucirc, 7636
\bibitem[Finn et~al.(1995)]{fin95} Finn,~R.~A., Fesen,~R.~A.,
    Darling,~C.~W., et~al. 1995, \apj, 110, 300
\bibitem[Fisher(2000)]{fis00} Fisher,~A. 2000, Ph.D. thesis, Univ. of Oklahoma
\bibitem[Franson, Lundqvist, \& Chevalier (1996)]{fra96} Franson,~C., Lundqvist,~P,
    \& Chevalier,~R.~A. 1996, \apj, 461, 993
\bibitem[Garbavich \& Kirshner(1997)]{gar97} Garnavich,~P., \& Kirshner,~R.
    1997, \iaucirc, 6529
\bibitem[Hachisu et~al.(1991)]{hac91} Hachisu,~I., Matsuda,~T.,
    Nomoto,~K., et~al. 1991, \apj, 368, L27
\bibitem[Harkness et~al.(1987)]{har87} Harkness,~R.~P., Wheeler,~J.~C.,
    Margon,~B., et~al. 1987, \apj, 317, 355
\bibitem[Hatano et~al.(1999a)]{hat99a} Hatano,~K., Branch,~D.,
    Fisher,~A., et~al. 1999, \apjs, 121, 233
\bibitem[Hatano et~al.(1999b)]{hat99b} Hatano,~K., Branch,~D.,
    Fisher,~A., et~al. 1999, \apj, 525, 881
\bibitem[Houck \& Frasson(1996)]{hou96} Houck,~J.~C., \&
    Fransson,~C. 1996, \apj, 456, 811
\bibitem[H\"{o}flich, Langer, \& Duschinger(1993)]{hof93} H\"{o}flich,~P.,
    Langer,~N., \& Duschinger,~M. 1993 \aap, 275, L29
\bibitem[Hill et~al.(1999)]{hil99} Hill,~G.~C., et al. 1999, \iaucirc, 7186
\bibitem[Iwamoto et~al.(1997)]{iwa97} Iwamoto,~K., Young,~T.~R., Nakasato,~N.,
    et~al. 1997, \apj, 477, 865
\bibitem[Jeffery \& Branch(1990)]{jef90} Jeffery,~D.~J, \& Branch,~D. 1990, in
    Jerusalem Winter School for Theoretical Physics, Vol.~6, Supernovae, ed.
    P.~Ruiz-Lapuente, R.~Canal, \& J.~Isern (Dordrecht: Kluwer), 659
\bibitem[Kifonidis et~al.(2000)]{kif00} Kifonidis,~K., Plewa,~T.,
    Janka,~H.~Th., et~al. 2000, \apj, 531, L123
\bibitem[Kurucz(1993)]{kur93} Kurucz,~R.~L. 1993, Kurucz CD-ROM, Atomic Data for Opacity Calculations
\bibitem[Lewis et~al.(1994)]{lew94} Lewis,~J.~R., Walton,~N.~A.,
    Meikle,~W.~P.~S., et~al. 1994, \mnras, 266, L27
\bibitem[Li \& McCray(1992)]{li92} Li,~H., \& McCray,~R. 1992,
    \apj, 387, 309
\bibitem[Li \& McCray(1993)]{li93} Li,~H., \& McCray,~R. 1993,
    \apj, 405, 730
\bibitem[Lucy(1991)]{luc91} Lucy,~L.~B 1991, \apj, 383, 308
\bibitem[Maeda et~al.(2000)]{mae00} Maeda,~K., Nakamura~T., Nomoto,~K.,
    et~al. 2000, \apjl, submitted (astro-ph/0011003)
\bibitem[Matheson et~al.(2000a)]{mat00a} Matheson,~T., Filippenko,~A.~V.,
    Barth,~A.~J., et~al. 2000, \apj, 120, 1487
\bibitem[Matheson et~al.(2000b)]{mat00b} Matheson,~T., Filippenko,~A.~V.,
    Ho,~L.~C., et~al. 2000, \apj, 120, 1499
\bibitem[Matheson et~al.(2001)]{mat01} Matheson,~T., Filippenko~A.~V.,
    Li,~W.~D., et~al. 2001, \aj, 121, 1648
\bibitem[Mazzali et~al.(1997)]{maz97} Mazzali,~P.~A., Chugai,~N.,
    Turatto,~M., et~al. 1997, \mnras, 284, 151
\bibitem[Mazzali, Iwamoto, \& Nomoto(2000)]{maz00} Mazzali,~P.~A., Iwamoto,~K.,
    \& Nomoto,~K. 2000, \apj, 545, 407
\bibitem[Millard et~al.(1999)]{mil99} Millard,~J., Branch,~D.,
    Baron,~E., et~al. 1999, \apj, 527, 746
\bibitem[Nagataki, Shimizu, \& Sato(1998)]{nag98} Nagataki,~S.,
    Shimizu,~T.~M., \& Sato,~K. 1998, \apj, 495, 413
\bibitem[Nakano \& Sumuto(1996)]{nak96} Nakano,~S., \& Akoi,~M. 1996, \iaucirc,
    6524
\bibitem[Nomoto et~al.(1993)]{nom93} Nomoto,~K., Suzuki,~T., Shigeyama,~T., et~al.
    1993, \nat, 364, 507
\bibitem[Nomoto, Iwamoto, \& Suzuki(1995)]{nom95} Nomoto,~K., Iwamoto,~K., \&
    Suzuki,~T. 1995, \physrep, 256, 173
\bibitem[Olson(1982)]{ols82} Olson,~G.~L. 1982, \apj, 255, 267
\bibitem[Prabhu et~al.(1995)]{pra95} Prabhu,~T.~P., Mayya,~Y.~D.,
    Singh,~K.~P., et~al. 1995, \aap, 295, 403
\bibitem[Podsiadlowski et al.(1993)]{pod93} Podsiaklowski,~Ph., Hsu,~J.~J.~L.,
    Joss,~P.~C., et~al. 1993, \nat, 364, 509
\bibitem[Rybicki \& Hummer(1978)]{ryb78} Rybicki,~G.~B, \&
    Hummer,~D.~G 1978, \apj, 219, 654
\bibitem[Saio, Kato, \& Nomoto(1988)]{sai88} Saio,~H., Kato,~M., \&
    Nomoto,~K. 1988, \apj, 331, 388
\bibitem[Schlegel, Finkbeiner, \& Davis(1998)]{sch98} Schlegel,~D.~J.,
    Finkbeiner,~D.~P., \& Davis, M. 1998, \apj, 500, 525
\bibitem[Shigeyama et~al.(1990)]{shi90} Shigeyama,~T., Nomoto,~K.,
    Tsujimoto,~T., et~al. 1990, \apj, 361, L23
\bibitem[Shigeyama et~al.(1994)]{shi94} Shigeyama,~T.,
    Suzuki,~T., Kumagai,~S., et~al. 1994, \apj, 420, 341
\bibitem[Sollerman, Leibundgut, \& Spyromilio(1998)]{sol98} Sollerman,~J.,
    Leibundgut,~B., \& Spyromilio,~J. 1998, \aap, 337, 207
\bibitem[Spyromilio(1994)]{spy94} Spyromilio,~T. 1994, \mnras,
    266, L61
\bibitem[Swartz et~al.(1993a)]{swa93a} Swartz,~D.~A.,
    Filippenko,~A.~V., Nomoto,~K., et~al. 1993, \apj, 411, 313
\bibitem[Swartz et~al.(1993b)]{swa93b} Swartz,~D.~A.,
    Clocchiatti,~A., Benjamin,~R., et~al. 1993, \nat, 365, 232
\bibitem[Turatto et~al.(1998)]{tur98} Turatto,~M., Mazzali,~P.~A.,
    Young,~T.~R., et~al. 1998, \apj, 498, L129
\bibitem[Patat, Chugai, \& Mazzali(1995)]{pat95} Patat,~E., Chugai,~N, \&
    Mazzali,~P.~A. 1995, \aap, 299, 715
\bibitem[Qiao et~al.(1996)]{qia96} Qiao,~Q.~Y., Li,~W.~D., Qiu,~Y.~L., et~al.
    1996, \iaucirc, 6527
\bibitem[Qiu et~al.(1999)]{qiu99} Qiu,~Y.~L., Li.~W.~D., Qiao,~Q.~Y., et~al.
    1999, \aj. 117, 736
\bibitem[Utrobin (1994)]{utr94} Utrobin,~V. 1994, \aap, 281, L89
\bibitem[Utrobin, Chugai, \& Adnronova(1995)]{utr95} Utrobin,~V., Chugai,~N.~N.,
    \& Adnronova,~A.~A. 1995, \aap, 295, 129
\bibitem[Utrobin (1996)]{utr96} Utrobin,~V.~P. 1996, \aap, 306, 219
\bibitem[Wang \& Hu(1994)]{wan94} Wang,~L.~F. \& Hu,~J.~Y. 1994, \nat, 369, 380
\bibitem[Wang \& Wheeler(1996)]{wan96} Wang,~L., \& Wheeler,~J.~C. 1996, \iaucirc,
    6351
\bibitem[Wang et~al.(2001)]{wan01} Wang,~L., Howell,~D.~A.,
    H\"{o}flich,~P., et~al. 2001, \apj, 550, 1030
\bibitem[Wheeler \& Filippenko(1996)]{whe96} Wheeler,~J.~C., \&
    Filippenko,~A.~V. 1996, in Supernova and Supernova Remnants, ed.
    R.~A.~McCray \& Z.~Wang (Cambridge University Press), 241
\bibitem[Williams(1987)]{wil87} Whilliams,~R.~E. 1987, \apj, 320,
    L117
\bibitem[Woosley et~al.(1994)]{woo94} Woosley,~S.~E.,
    Eastman,~R.~G., Weaver,~T.~A., et~al. 1994, \apj, 429, 300
\bibitem[Young, Baron, \& Branch(1995)]{you95} Young,~T.~R., Baron,~E., \&
    Branch,~D. 1995, \apj, 449, L51
\bibitem[Zhang et~al.(1995)]{zha95} Zhang,~Q., Hu,~J.~Y.,
    Wang,~L.~F, et~al. 1995, \mnras, 277, 1115
\bibitem[Zhang \& Wang(1996)]{zha96} Zhang,~Q., \& Wang,~Z.~R.
    1996, \aap, 307, 166

\end{thebibliography}
\end{document}